# Topology-Aware Spatio-Temporal Graph Transformer for Predicting Smart Grid Failures

Anh Le, PhD Student, North Dakota State University

Phat K. Huynh, PhD, North Carolina A&T State University

Om P. Yadav, PhD, North Carolina A&T State University

Harun Pirim, PhD, North Dakota State University

Chau Le, PhD, University of North Carolina at Charlotte

Trung Q. Le, PhD, University of South Florida

*SUMMARY & CONCLUSIONS*

Smart grid infrastructure needs improved resilience and preventive maintenance through more accurate predictions. Current methodologies lack accurate representation of spatio-temporal-causal interdependencies and class imbalance in failure prediction tasks. This study introduces a Topology-Aware Spatio-Temporal Graph Transformer (ST-GT) architecture that overcomes existing limitations by using three main innovations: (1) directly incorporating physical transmission network topology into the transformer attention mechanism to identify spatial failure propagation patterns; (2) unified processing of static topological descriptors and (3) temporal Phasor Measurement Units (PMU) sequences in an end-to-end framework.

The ST-GT model exhibited outstanding performance in five-fold cross-validation across 10 substations, attaining perfect recall ($1.000 \pm 0.001$) and an F1-score of $0.858 \pm 0.009$, markedly surpassing XGBoost baselines (0.683 accuracy/F1). Perfect recall guarantees that no critical failures are overlooked, which is essential for grid safety; however, it may lead to an increase in false alarm rates. This framework integrates temporal dynamics modeling with spatial graph awareness for critical infrastructure monitoring. It offers interpretable insights into failure propagation pathways and enhances maintenance strategies.

Future research focuses on developing cost-weighted loss functions for precision-recall trade-off enhancement, implementing real-time monitoring systems with uncertainty quantification, and creating cost-sensitive frameworks balancing false alarm expenditures with failure consequences. The methodology's success suggests its potential for wider application in critical infrastructure areas requiring spatio-temporal failure prediction.

## 1. INTRODUCTION

Smart grid vulnerabilities from cyberattacks, equipment failures, and weather events have tripled since the 1980s [1] [2]. Outage data analysis is foundational for grid reliability and predictive maintenance [3]. Outage information comes from electric companies and includes event timestamps, location identifiers, failure causes, and equipment types[4, 5]. Detailed outage data is foundational for various critical aspects of power grid management and research: reliability and resilience [3], predictive maintenance [6], vulnerability assessment [1].

The literature on power grid analysis utilizes three main methodologies to tackle predictive maintenance, resilience, and outage detection: Statistical and probabilistic modeling quantifies risk and predicts events using historical data. Machine learning and AI approaches leverage data-driven algorithms for predictions and classifications. Network theory and graph-based approaches model grid structures to understand behavior and vulnerabilities.

- Statistical and probabilistic modeling is commonly used to measure risks, predict occurrences, and evaluate power system reliability [7]. Empirical formulas and probability distributions like Weibull and Poisson are used to model failure rates and time between failures [7]. Markov Considerations of system states and transitions are used to measure reliability [8]. For probabilistic risk and vulnerability assessment, Self-Organized Criticality theory analyzes power-law outage size distributions to uncover catastrophic failure warning signals for probabilistic risk and vulnerability assessment [4]. Bayesian techniques anticipate outages using prior distributions and likelihood functions, while Generalized Linear Models can create failure rate models from weather data [5]. However, these methods face limitations including the assumption of stationary failure distributions and difficulty in modeling complex temporal dependencies and cascading effects [7]. In addition, traditional statistical models may not capture the dynamic, cascading nature of failures or presume full temporal visibility of incident data [6].

- Machine learning techniques employ data-driven algorithms for outage prediction and grid analysis, comprising three main methodologies: traditional ensemble methods, deep learning architectures, and graph neural networks.

  Traditional ensemble methods combine multiple models to enhance predictive performance, including AdaBoost, stacking ensemble learning, XGBoost, and Random Forest for predicting daily reported incidents in electrical distribution networks [9]. Ghasemkhani et al. (2024) [3] achieved 98.4% accuracy using XGBoost with Maximum Relevance Minimum Redundancy feature selection from PMU data. However, these models extract features from temporal, geographic, and graph data using discrete rather than sequential processing, failing to capture temporal and spatial correlations essential for comprehensive grid analysis.

  Deep learning architectures address sequential data limitations through Recurrent Neural Networks (RNNs), Gated Recurrent Units (GRUs), and Long Short-Term Memory (LSTM) networks. Branco et al. [7] integrated wavelet packet denoising with LSTM networks for daily fault prediction, surpassing conventional ensemble approaches with reduced root mean square error. Prieto Godino et al. (2025) [10] created diverse deep learning ensembles (RNN, GRU, LSTM, and Convolutional Neural Network) trained on weather-related outage data. Even though temporal modeling has gotten more effective, there are still significant challenges. For example, using datasets from a univariate time series makes it hard to apply to other situations. Furthermore, spatial relationships are not considered, and network topology is not integrated.

  Graph neural networks (GNN) demonstrate a wide range of capabilities in enhancing power grid management and resilience [6]. The multilayer GNN framework achieved a 30-day high F1-score for predictive maintenance. A novel Dual-Graph Structure GNN (DGS-GNN) has also been applied to real-time topology detection, which transforms the problem into an inductive link prediction task without requiring prior knowledge of network parameters [11]. However, their GNN implementations still treat substations as separate entities, and traditional methods do not consider how failures can spread through spatial, temporal, and causal links. This leads to fragmented resilience strategies that do not fully understand how power grid operations are connected.

- Network-dynamics modeling in power grids focuses on understanding and predicting the temporal evolution and dynamic responses to events such as cascading failures and blackouts [3, 12-14]. The analysis of historical blackout size distributions shows hefty tails or power laws, showing that while huge incidents are rare, they are expected [14]. Topology-aware modeling integrates the grid's structural information, essential for tasks such as topology detection, state estimation, and vulnerability analysis [15]. Advanced methods, specifically GNNs and transformer models, integrate these two domains by utilizing the graph-like characteristics of power grids and concurrently handling dynamic and time-series data. The models are intended to capture intricate interdependencies across spatial, temporal, and causal dimensions, thereby improving predictive maintenance, resilience clustering, and the identification of vulnerable components or attack paths, with the goal of achieving proactive, interpretable, and scalable grid management. However, these approaches face limitations from data scarcity, particularly for rare cascading events, which reduces model generalizability.

This study presents a Topology-Aware Spatio-Temporal This research introduces the Spatio-Temporal Graph Transformer (ST-GT), a deep learning framework designed for forecasting next-day substation failures in smart grid systems. Existing methodologies face challenges in effectively capturing spatial-temporal-causal interdependencies and cascading failures. ST-GT mitigates these limitations via three primary innovations: (1) Integrating physical network topology into the attention mechanism, (2) combining static network descriptors with temporal dynamics within an end-to-end architecture, and (3) addressing significant class imbalance in failure datasets.

## 2. *METHODOLOGY*

We introduce ST-GT (Figure 1), a comprehensive transformer framework designed for the prediction of next-day substation failures.

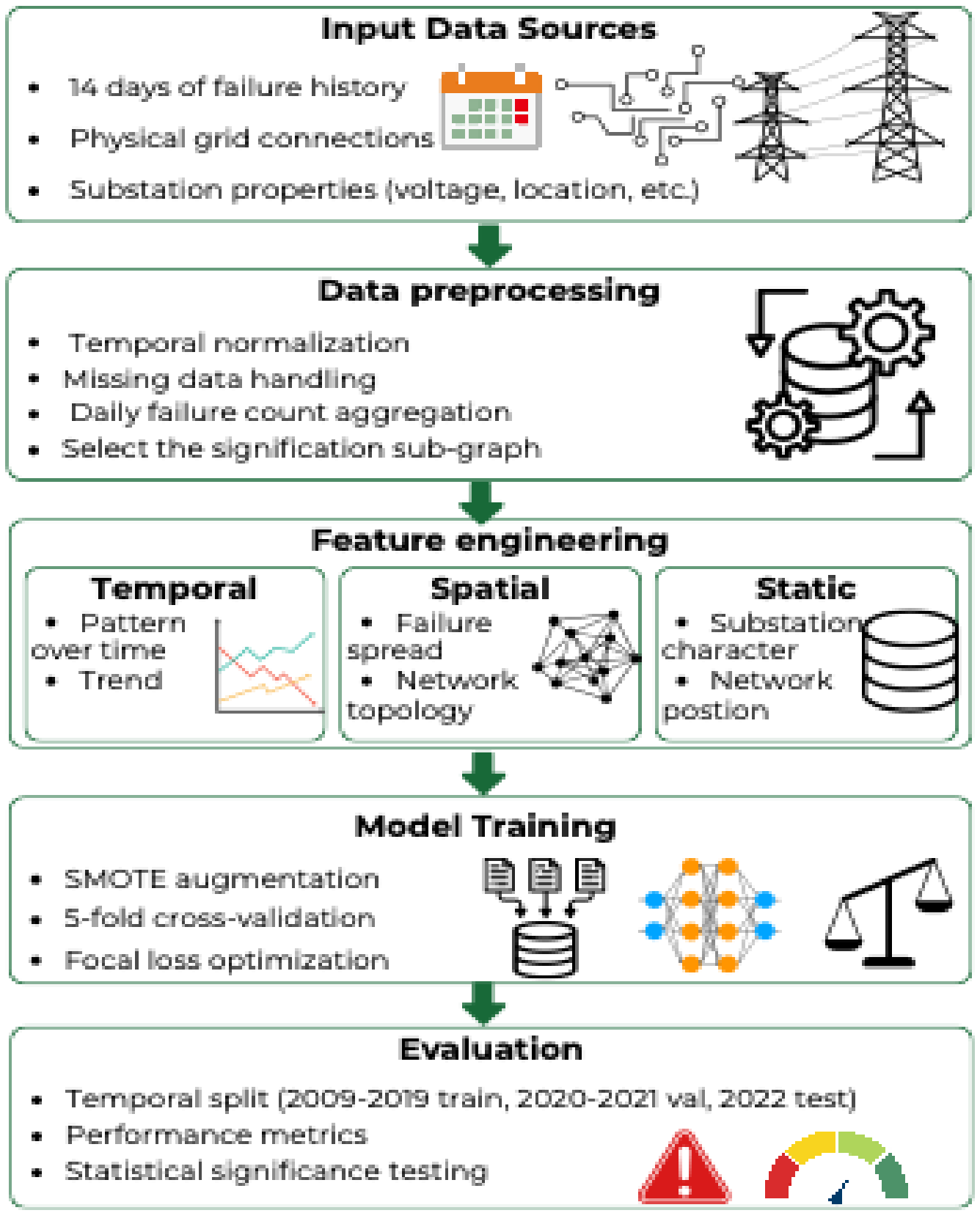

*Figure 1. Methodology overview*

The model integrates 14-day historical failure data with static network characteristics, employing self-attention mechanisms to discern temporal failure patterns. A graph-based attention layer utilizing an adjacency matrix effectively captures spatial failure propagation, and the resultant embeddings are integrated with static data for classification through a three-layer MLP trained using focal loss. The model undergoes validation by five-fold cross-validation, incorporating Synthetic Minority Oversampling Technique (SMOTE) augmentation and bootstrap-based feature selection.

*2.1 Problem Formulation and Mathematical Notation*

We formalize the substation failure prediction problem as a spatio-temporal binary classification task. Let $S = 1,2,\dots N$ denote the set of $N$ substations in the electrical grid, substation $s \in S$ is characterized by both temporal failure patterns and static network properties. We define $d_{s,t} \in N_0$ as the daily failure count recorded at substation $s$ on day $t$, and where $N_0$ denotes the set of non-negative integers. The binary prediction target is expressed as: $y_{s,t} = 1[d_{s,t} > 0] \in \{0,1\}$ where $\mathbf{1}[\cdot]$ represents the indicator function that equals 1 when the condition is satisfied and 0 otherwise.

Each substation '$s$' possesses a static attribute vector $z_s \in R^{F_z}$ containing $F_z$ engineered features derived from network topology, geographical properties, and operational characteristics. The physical connectivity of the transmission network is represented by an undirected graph $\mathcal{G} = (S, E)$, where $E$ denotes the set of transmission lines connecting substations, and the corresponding adjacency matrix $A \in \{0,1\}^{N\times N}$ encodes the network topology.

Given a look-back horizon of $L$ days (set to L = 14 based on domain expertise and preliminary experiments), we construct temporal windows as:

$$\boldsymbol{X_{s,t}} = [d_{s,t-L+1},\ d_{s,t-L+2},\ \dots,\ d_{s,t}] \in R^L \tag{1}$$

The ST-GT model learns a parametric mapping function: $f_\theta: (X_{s,t},\ z_s,\ A) \longrightarrow \hat{p}_{s,t+1}$ where $\theta$ represents the learnable parameters, and $\hat{p}_{s,t+1} \in (0,1)$ denotes the predicted probability of failure occurrence at substation $s$ on day $t + 1$.

*2.2 Data Acquisition and Preprocessing Pipeline*

Data on historical failure incidents from 533 substations have been extracted from the PMU history database of Oklahoma Gas and Electricity, encompassing the timeframe from January 2009 to August 2022. The preprocessing pipeline normalizes timestamps, links events with substations, and consolidates occurrences into daily failure counts $d_{s,t}$ for each substation-day pairing. Days lacking observations are allocated zero failure counts to maintain temporal consistency. Substations with fewer than 180 observation days or fewer than 100 total failure incidents are excluded to ensure adequate statistical power. The finalized collection maintains a strict chronological sequence to prevent temporal leakage and encompasses substations within the Oklahoma smart grid network, along with comprehensive operational metadata.

*2.3 Network Topology Construction and Analysis*

*2.3.1 Geographical Proximity Graph*

Network topology encompasses both physical transmission lines and spatial configurations. We create a proximity-based adjacency matrix that designates geographically adjacent substations as related:

$$A_{ij} = A_{ji} = \begin{cases} 1, & if\ dist_{gc}(i,j) < \tau \\ 0, & otherwiise \end{cases} \tag{2}$$

where $dist_{gc}(i,j)$ is the great-circle distance between substations $i$ and $j$, whereas τ represents the proximity threshold. Due to common environmental elements, cascade effects, and regional grid stress, disruptions can influence surrounding substations, hence this spatial method depicts failure propagation patterns beyond direct electrical connectivity.

*2.3.2 Topological Feature Extraction*

We compute graph-theoretic properties of each substation's topological role from the network. Table 1 presents the mathematical formulations for each topological feature, where the notation is defined as follows: $\sigma_{ij}(s)$ represents the number of shortest paths between nodes i and j that pass through node s, $\sigma_{ij}$ is the total number of shortest paths between nodes i and j, d(s,j) is the shortest path distance between nodes s and j, N(s) denotes the set of neighbors of node s, $e_{jk}$ represents an edge between nodes j and k, and $v_j$, $v_k$ represent vertices j and k respectively.

*Table 1. Topological feature formula*

| Features | Formula |
|---|---|
| Degree centrality ($C_D$) | $C_D(s) = \sum_{j=1}^{N} A_{sj}$, |
| Betweenness Centrality ($C_D$) | $C_B(s) = \sum_{i\neq j} \frac{\sigma_{ij}(s)}{\sigma_{ij}}$ |
| Closeness Centrality ($C_C$) | $C_C(s) = \frac{1}{\sum_{j=1}^{N} d(s,j)}$ |
| PageRank (PR) $\alpha = 0.85$ | $PR(s) = \frac{1-\alpha}{N} + \alpha \sum_{j\in N(s)} \frac{PR(j)}{\lvert N(j)\rvert}$ |
| Clustering Coefficient (CC) | $CC(s) = \frac{2\lvert\{e_{jk}: v_j, v_k \in N(s), e_{jk} \in E\}\rvert}{\lvert N(s)\rvert(\lvert N(s)\rvert - 1)}$ |

*2.4 Feature Engineering and Selection*

Static Feature Selection: We utilize bootstrap aggregation through 100 random forest iterations, top_feature=15, stability_threshold=0.8 to ensure robust selection [9]. The selection process involves identifying the 15 features exhibiting the lowest coefficient of variation across bootstrap iterations, thereby ensuring stable predictive signals and optimizing computational efficiency.

Temporal Feature Processing involves the use of standard seasonal encodings, including sinusoidal representations for day-of-year, one-hot encoding for weekdays and months, linear time counters, and weekend indicators. Additionally, a log-transformation, log(d + 1), is applied to failure counts to stabilize variance while maintaining the integrity of zero-count data. Raw failure counts undergo log-transformation to reduce the impact of extreme values: $\tilde{d}_{s,t} = \log(1 + d_{s,t})$.

Class imbalance and normalization: We propose a three-stage augmentation strategy with a failure rate of less than 5%. Initially, we replicate positive samples by applying Gaussian noise ($\sigma^2 = 0.05^2$). For oversampling, we generate synthetic samples using SMOTE with five nearest neighbors [1], [16]. We admit that SMOTE augmentation may produce synthetic correlations not in the original data. We (1) employ SMOTE only to training sets during cross-validation, (2) ensure temporal separation between training and testing data to prevent information leaking, and (3) validate our conclusions on 2022 test data without fake samples to reduce this bias. Lastly, we proceed until failure days constitute 30% of the dataset. This ensures realistic deployment conditions and offers adequate positive instances.

RobustScaler is applied to temporal data to ensure outlier resistance, while StandardScaler is utilized for static features to achieve Gaussian normalization. This approach results in an integrated preprocessing pipeline that improves convergence and maintains the characteristics of feature categories.

### *2.5 ST-GT Architecture Design*

The ST-GT architecture combines temporal pattern recognition with spatial relationship modeling using a meticulously crafted transformer-based framework. The model analyzes temporal sequences for specific substations while concurrently accounting for the network-wide dissemination of failure occurrences through topology-aware attention methods.

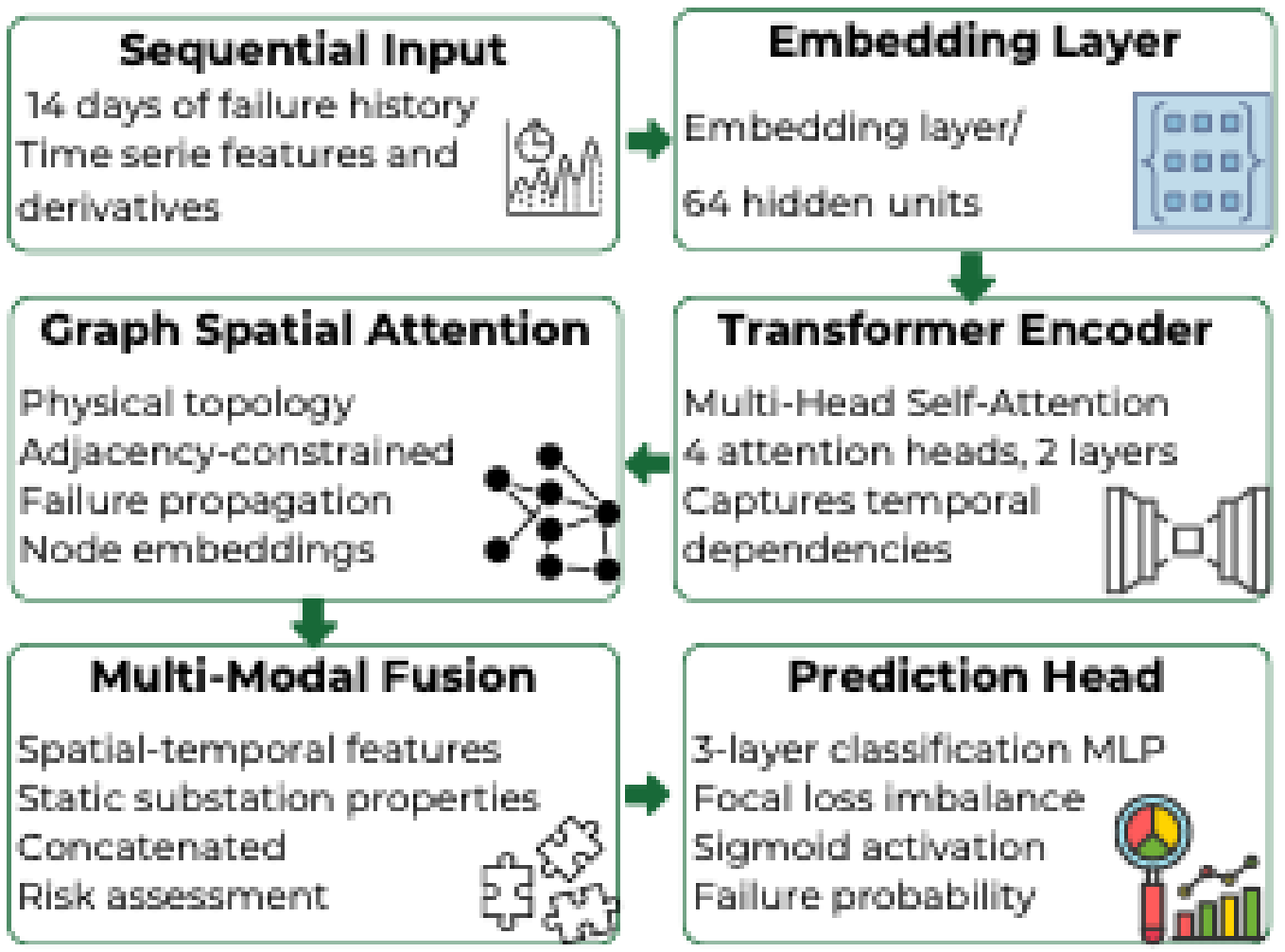

*Figure 2. Spatio-Temporal Graph Transformer design*

#### *2.5.1 Input Embedding Layer*

The raw input features $x_k \in R^{F_x}$ (which includes failure counts and temporal features) are projected onto a higher-dimensional embedding space for each day $k$ in the temporal window of length L:

$$fh_k = W_x \boldsymbol{x_k} + \boldsymbol{p_k} \tag{3}$$

where $W_x \in R^{d \times F_x}$ is a learnable linear transformation matrix with embedding dimension $d$ = *64*, and $p_k \in R^d$ represents learnable positional encodings to the embedding dimension $d$ = *64* balances model expressiveness and computational efficiency, capturing intricate temporal patterns while preserving appropriate memory requirements for graph processing.

#### *2.5.2 Data Processing Pipeline*

This subsection outlines three parallel processing components that convert raw inputs into representations suitable for modeling: temporal encoding, topology-aware spatial attention, and static feature processing (Figure 1).

The ST-GT architecture concurrently processes three data streams to produce representations suitable for modeling. The embedded temporal sequence ($H = [h_1, h_2, \dots, h_L] \in R^{L \times d}$) undergoes processing via two successive transformer encoder blocks utilizing multi-head self-attention.

$$\mathrm{MultiHead(H)} = \mathrm{Concat}(\mathrm{head}_1, \dots, \mathrm{head}_h)\, \mathrm{W}^{\mathrm{O}} \tag{4}$$

Each attention head $i$ computes: $head_i = Attention\left(HW_i^Q, HW_i^K, HW_i^V\right)$ Learnable projection matrices $W_i^Q, W_i^K, W_i^V \in R^{d \times d_k}$ are utilized for $h = 8$ attention heads. The final temporal representation $h^* = \widetilde{H_L} \in R^d$ encapsulates the temporal dynamics inherent in the failure sequence of each substation.

Secondly, our main innovation incorporates transmission network topology through spatial attention. Each substation $s$ possesses a learnable node embedding ($e_s \in R^d$). The spatial attention components queries, keys, and values are formulated as follows:($q_s = k_s = v_s = h^* + e_s$). For mini-batches comprising B substations, the topology-constrained attention is calculated as:

$$SpatialAttn(Q, K, V) = \left[softmax\left(\frac{QK^\top}{\sqrt{d}}\right) \odot A_B\right] V \tag{5}$$

Let $A_B \in \{0,1\}^{B \times B}$ denote the adjacency sub-matrix, and let $\odot$ represent element-wise multiplication, thereby restricting attention to physically connected substations.

Third, static features ($z_s \in R^{F_z}$) are processed using a specialized two-layer MLP: The equation $static(z_s) = ReLU(W_1 z_s + b_1) \cdot W_2 + b_2$ produces an output dimension of $d/2 = 32$, resulting in processed static representations $z_s^* \in R^{d/2}$ that encapsulate time-invariant characteristics of substations. The three processed representations are prepared for feature fusion and subsequent classification.

#### *2.5.3 Final Prediction and Model Optimization*

This subsection includes three concluding processing steps: multi-modal feature fusion, classification head prediction, and focal loss optimization for addressing class imbalance.

The spatially aware temporal representations $g_s$ and processed static features $z_s^*$ are combined to create a comprehensive feature vector: $f_s = [g_s; z_s] \in R^{3d/2}$.

encompassing temporal dynamics, spatial relationships, and static characteristics. The fused feature vector $f_s$ undergoes processing via a three-layer classification head to yield the final failure probability.

$$ls, t = W_3 \cdot \text{ReLU}(W_2 \cdot \text{ReLU}(W_1 \mathbf{f}s + b_1) + b_2) + b_3 \quad (6)$$

The output logit ($l_{s,t} \in R$) is transformed into probability using the sigmoid function. Let $\hat{p}_{s,t+1}$ be defined as $\sigma(l_{s,t})$, which is expressed mathematically as $\hat{p}_{s,t+1} = \frac{1}{1+\exp(-l_{s,t})}$.

To mitigate extreme class imbalance, we utilize focal loss, which reduces the weight of well-classified instances.

$$\mathcal{L} = \beta \cdot (-\alpha(1 - p_t)^\gamma \log(p_t)) \quad (7)$$

Let $p_t = p$ when $y = 1$; otherwise, let $p_t = 1 - p$. Hyperparameters are established as $\alpha = 0.3$ and $\gamma = 2.0$, with class-frequency weighting defined as $\beta = (1 - r)/r$, where $r$ denotes the global positive class ratio, thereby ensuring adequate focus on infrequent failure events.

*2.6 Training and Evaluation Method*

The ST-GT model uses 512-mini-batches, weighted sampling, mixed precision training, gradient clipping, and early stopping for stable convergence. We employ a temporal cross-validation approach wherein:

- The data is divided temporally (training: 2009-2019, validation: 2020-2021, testing: 2022)
- During the training period, 5-fold validation guarantees the absence of temporal leakage.
- Test substations are physically isolated to assess spatial generalization.

Since missing failures costs more than false alarms, recall-weighted F1-score prioritizes sensitivity when setting decision thresholds. Bootstrap resampling with 1,000 iterations for 95% confidence intervals determines statistical significance. Implementation was conducted in Python 3.9 using PyTorch 1.12 for the transformer architecture, scikit-learn 1.1 for preprocessing and baseline models, and PyTorch Geometric 2.1 for graph operations. The NVIDIA A100 GPU with 40 Gb efficiently performs five-fold cross-validation across three years of data for 15 substations.

## 3. RESULT

The ST-GT model performed evaluation through five-fold cross-validation across 10 substations, exhibiting substantial enhancements in failure prediction performance compared to baseline techniques. Our investigation was conducted in three principal phases: network topology characterization, feature importance assessment, and comparative evaluation of model performance.

Initially, we defined the spatial distribution of failure incidents throughout the power grid network to figure out failure patterns and propagation routes. The investigation indicated that disturbances displayed spatial clustering, with specific substations encountering markedly greater failure rates than others. The examination of network topology revealed the spatial clustering of failure-prone locations and the potential propagation of disturbances through the network infrastructure via directional connections. The findings validated the significance of integrating topological information into our prediction approach, as demonstrated in Figure 3.

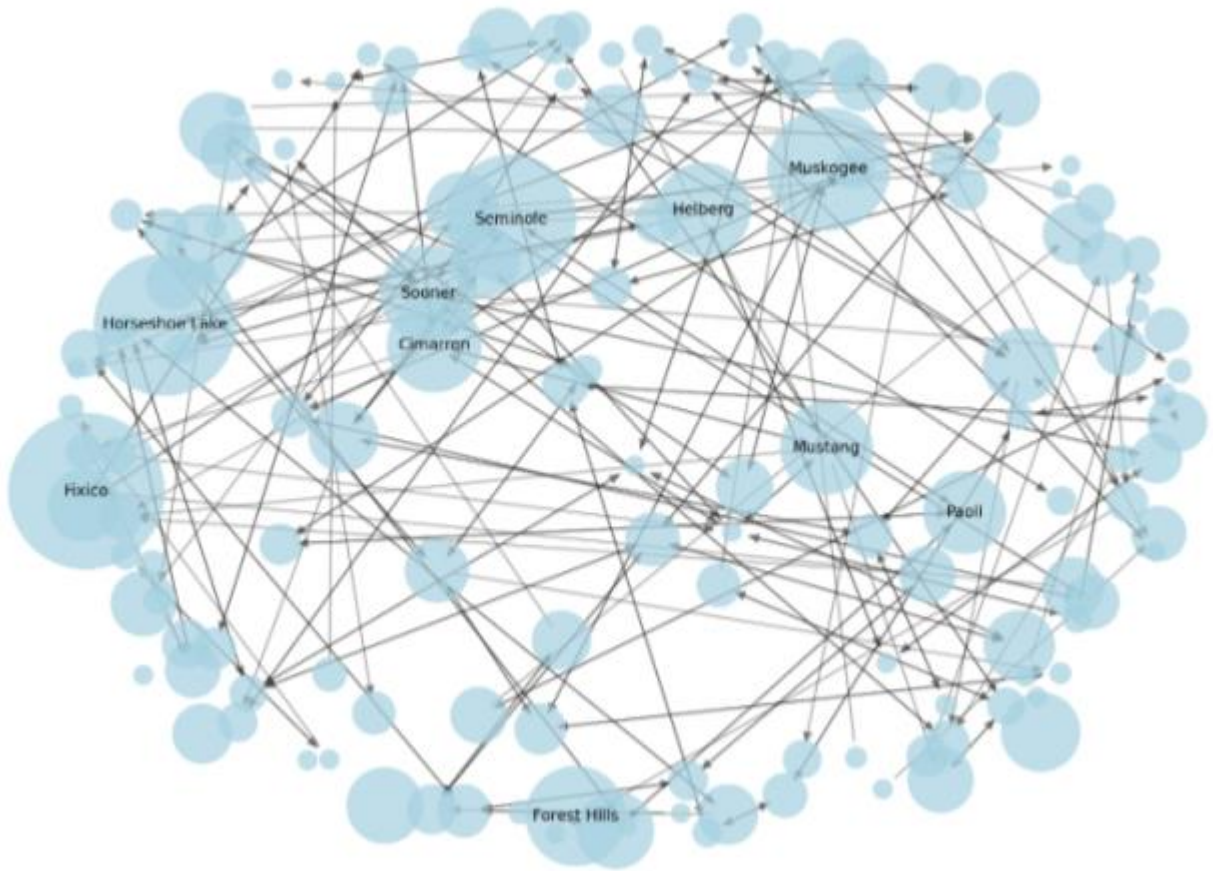

*Figure 3. Network topology: disturbance count by nodes*

We performed feature significance analysis across many feature categories to determine the most significant predictors of failure events. The investigation indicated that voltage characteristics (V69) and operational metrics (Count—number of connections/links surrounding substations) were the most critical predictors of failure. This research demonstrated that indications of power system stress exerted greater influence than assessments of network structure. The dominance of electrical factors over spatial characteristics indicated that the system's operating status was more significant than geographical location in predicting failures, as illustrated in Figure 4.

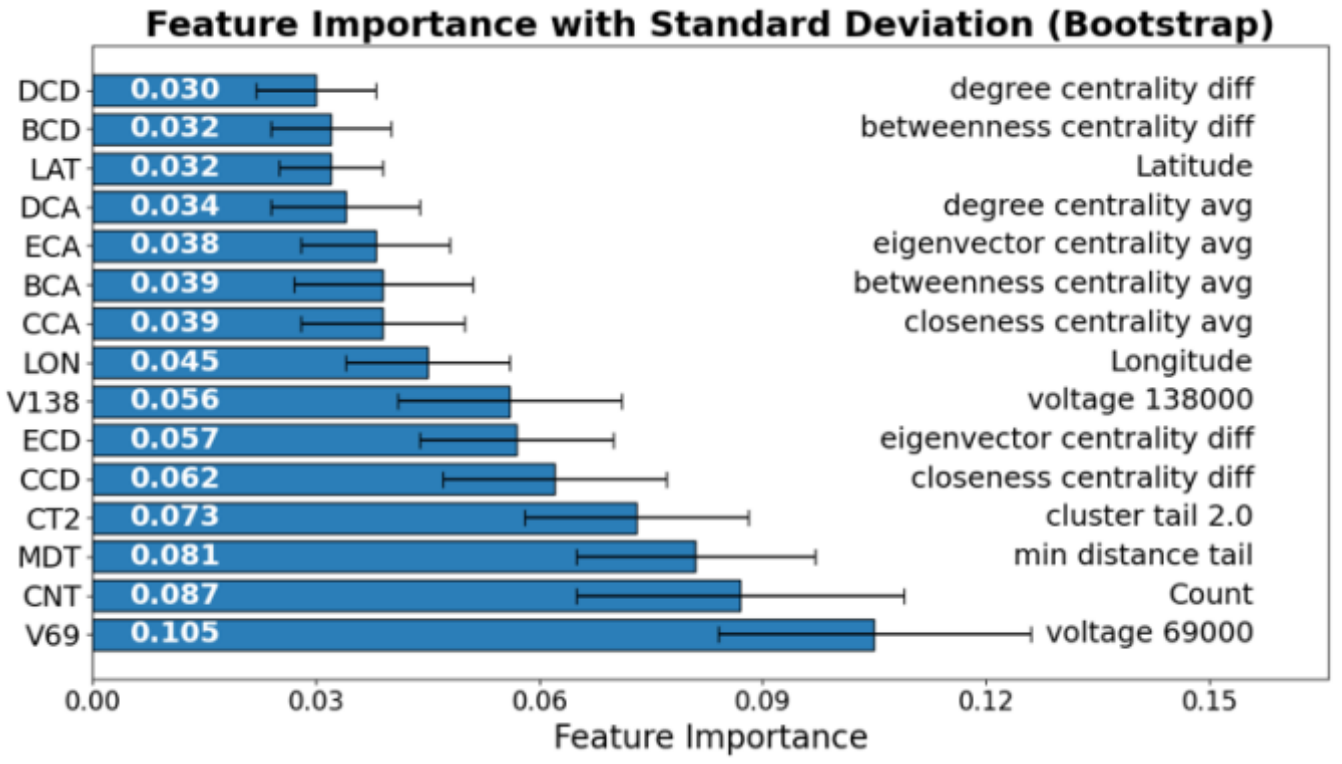

*Figure 4. Feature importance*

We assessed model performance using both univariate and multivariate analyses. The univariate analysis employing 5-step sequences across several feature sets demonstrated differing performance levels, with ensemble approaches attaining the minimal MAE values (Table 2). Nevertheless, the near-zero MAE numbers may be deceptive in situations of extreme class imbalance, as a always-zero predictor would yield comparable accuracy.

*Table 2. Mean absolute error for Section 80, 5-step sequences*

| Feature set | LSTM | GRU | CNN | Ensemble |
|---|---|---|---|---|
| temporal _cause | 0.026 | 0.014 | 0.001 | 0.001 |
| temporal _spatial | 0.007 | 0.004 | 0.007 | 0.003 |
| temporal _only | 0.005 | 0.010 | 0.121 | 0.005 |
| temporal _topology | 0.082 | 0.004 | 0.029 | 0.006 |
| all _features | 0.008 | 0.078 | 0.019 | 0.010 |

The multivariate analysis of XGBoost baselines and ST-GT across 10 substations demonstrated significant performance disparities. Hyperparameters were optimized through grid search: max_depth=6, learning_rate=0.1, n_estimators=100, subsample=0.8. The same focal loss function and class weighting strategies were applied to ensure fair comparison. XGBoost demonstrated consistent albeit constrained performance (~0.683) across many feature combinations, irrespective of the inclusion of temporal, spatial, topological, or causal features. Conversely, ST-GT exhibited remarkable performance, attaining an accuracy of 0.750 ± 0.014, a precision of 0.750 ± 0.014, flawless recall (1.000 ± 0.001), and an F1-score of 0.858 ± 0.009 (Table 3).

*Table 3. Multivariate performance comparison across 10 substations: XGBoost vs. ST-GT*

| Model | Accuracy (±std) | Precision (±std) | Recall (±std) | F1-Score (±std) |
|---|---|---|---|---|
| Temporal Cause Spatial Topology | 0.683 ± 0.012 | 0.683 ± 0.013 | 0.683 ± 0.012 | 0.683 ± 0.012 |
| Temporal Spatial | 0.683 ± 0.013 | 0.684 ± 0.013 | 0.683 ± 0.013 | 0.683 ± 0.013 |
| Temporal Topology | 0.683 ± 0.014 | 0.683 ± 0.014 | 0.683 ± 0.014 | 0.683 ± 0.014 |
| Temporal Cause | 0.682 ± 0.015 | 0.682 ± 0.015 | 0.682 ± 0.015 | 0.682 ± 0.015 |
| Temporal | 0.681 ± 0.015 | 0.681 ± 0.015 | 0.681 ± 0.015 | 0.681 ± 0.015 |
| **ST-GT** | **0.750 ± 0.014** | **0.750 ± 0.014** | **1.000 ± 0.001** | **0.858 ± 0.009** |

The exceptional recall illustrates our paramount design priority for safety, as overlooking failures might lead to catastrophic consequences. Although the application of focal loss, the severe class imbalance required prioritizing failure detection over the minimization of false alarms.

## 4. *CONCLUSION*

A Topology-Aware Spatio-Temporal Graph Transformer (ST-GT) was introduced for forecasting next-day failure incidents across all substations within a smart grid. The three primary innovations are: (1) Incorporating the physical transmission network directly into the transformer's attention mechanism; (2) Merging static topological descriptors with temporal PMU sequences into a unified end-to-end model; (3) Tackling class imbalance via targeted augmentation and the implementation of focal loss. In five-fold cross-validation, ST-GT attained flawless recall (1.000 ± 0.001) and an F1-score of 0.858 ± 0.009, markedly surpassing the XGBoost baseline (0.683 accuracy/F1). Perfect recall (1.000 ± 0.001) is a common outcome with highly imbalanced datasets (4.97% positive class rate) when prioritizing safety over precision. We deliberately designed our model to minimize missed failures rather than false alarms. Cross-validation on temporally separated data confirms robust generalization rather than overfitting.

## *ACKNOWLEDGEMENTS*

This research is partially supported by National Science Foundation (NSF) EPSCoR RII Track-2 Program under the grant number OIA-2119691 at North Dakota State University.

## *BIOGRAPHIES*

Anh Le, PhD Student
Department of Civil Engineering North Dakota State University
e-mail: ducanh.le@ndsu.edu

Anh Le is currently a PhD Student of the Department of Civil Engineering at North Dakota State University. His research interests focus on advanced deep learning in smart energy systems, computational fluid dynamics.

Phat K. Huynh, PhD
Department of Industrial and Systems Engineering
North Carolina A&T State University
e-mail: pkhuynh@ncat.edu

Dr. Phat K. Huynh is currently an Assistant Professor of the Department of Industrial and Systems Engineering at North Carolina A&T State University. His work focuses on complex systems modeling and predictive analytics in healthcare.

Om Prakash Yadav, PhD
Department of Industrial and Systems Engineering
North Carolina A&T State University
e-mail: oyadav@ncat.edu

Dr. Om Prakash Yadav is currently a Professor and Chair of the Department of Industrial and Systems Engineering at North Carolina A&T State University. His research interests include reliability modeling and analysis, risk assessment, robust design optimization, and manufacturing systems analysis.

Chau Le, PhD
Department of Engineering Technology & Construction Management
University of North Carolina at Charlotte
e-mail: cle12@charlotte.edu

Dr. Chau Le is currently an Assistant Professor in the Department of Engineering Technology & Construction Management at University of North Carolina at Charlotte. His research interests include human safety and health, construction and infrastructure management, and artificial intelligence.

Harun Pirim, PhD
Department of Industrial and Manufacturing Engineering
North Dakota State University
e-mail: harun.pirim@ndsu.edu

Dr. Harun Pirim is currently an Assistant Professor of the Department of Industrial and Manufacturing Engineering at North Dakota State University. His research interests focus on network science, mathematical programming, and machine learning in biological, social, and decision sciences.

Trung (Tim) Q. Le, PhD
Department of Industrial & Management Systems Engineering
University of South Florida
e-mail: tqle@usf.edu

Dr. Trung (Tim) Q. Le is currently an Assistant Professor of the Industrial and Management Systems Engineering Department at the University of South Florida. His research focuses in 3 main directions: 1) data-driven and sensor-based modeling, 2) medical device manufacturing and bio-signal processing, and 3) predictive analytics for personalized healthcare.